# Shaping the international financial system in century of globalization.

V.O. Ledenyov and D.O. Ledenyov

*Abstract* - We educe a perspective on how best to regulate the bank of tomorrow in frames of debate launched by the International Centre for Financial Regulation and Financial Times. Our goal is to create a conceptual framework for policymakers and regulators to shape the international financial system in century of globalization using the 1888 FT's motto: "Without fear and without favour." Our prospect employs an analytical approach, which focuses on the origins and evolution of banking system, its transformation over the recent decades, subsequent encountering the limits to growth and redefinition of new strategic boundaries of emerging financial industry. We identify the main reasons and limitations, which led to the global financial crisis. We propose the new research agendas with the aim to understand the situation in finances, evaluate the created systemic damages, and find the possible ways to resolve the existing problems through introduction of new banking regulation. We think that the global economic and financial systems are highly nonlinear systems. In our opinion, the frequency, phase and amplitude modulation during the mixing of waves, which characterize the Kitchin, Juglar, Kuznets, Kondratiev economic cycles, may result in origination of strong nonlinear dynamics in financial system, accompanied by chaos-induced phenomena. These nonlinear effects have to be taken to the account, when adding the liquidity to the financial system in small quantas in series over time period during the Quantitative Easing policy execution by central banks. We propose the Random Tax to be selectively imposed on the profits, obtained by market agents during high-risk high-profit speculative transactions. We expect that the Random Tax will stabilize the financial system in conditions of free market capitalism. We conclude by outlining a new set of policies to regulate the financial system toward its sustainable development in harmony with global society and planetary ecosystem.

For many centuries, people were dreaming about finding the perfect solutions for imperfect problems, which they are about to face in real World. Considering the evolution of financial industry over the resent decades, one question may arise in Gavetti, Rivkin (2007): Where does a bank's competitive strategy come from? It is obvious that the wisdom of policy makers in creation of banking regulation and monetary policies as well as the skills and experiences by bankers in search for virtuous strategies toward the profitable and sustainable bank operation, and many other internal and external (to the financial industry) factors and forces may have a great impact on the origin of bank's competitive strategy in given economic conditions. Despite of considerable progress in innovation introduction in financial industry over the recent years, our global society remains weak in search of meaningful answer to a set of straightforward questions: Why did the present monetary policies lead the financial industry to the crisis? How would it be possible for the systemic regulators to resolve the existing problems and avoid the repetition of up and down cycles in financial industry in the future? How can the financial industry continue to serve the interests of investors and economic growth without putting our World at risk in Chaney, Goodhart, Webb (2009)?

Let us go back to basics and try to understand the right meanings of definitions and terminology used in the field of finances in Hirsch (1896): "The primary object of a Bank is the purchase of debts with its own credit. Credit is the exchange of existing commodities at a future date, which later may already be in existence, or may be produced in the future. The economic object of credit is to enable producers to obtain capital. Capital comprises all wealth produced for the ultimate purpose of satisfying some want or desire, but actually employed in adding to the productiveness of future labor. Every transfer of capital on credit creates a Debt, i.e. a legal right to other commodities at a future date."

In general, the bank is a maturity and risk transforming institution in Sargent (2010). There are two investment strategies used by banks presently in Flassbeck (2009):

1) To invest with the goal to build the productive capacities;

2) To invest in the Structured Investment Vehicles with the aim to trade the derivatives in financial and/or equity markets.

The Structured Investment Vehicles are extremely complex. The purpose of structured credit products is to give fixed income investors fully rated and leveraged exposure to the main Credit Derivatives Indices. Investment Portfolio building with positions hedged by different means depends on investor's expertise in the following investment vehicles:

1. Credit Derivatives (CD are the financial contracts whose payoffs explicitly depend on the behaviour of one or more indices)**:** Collateralized Debt Obligations (CDO), Constant Proportion Debt Obligations (CPDO), and investment protection mechanisms such as the Synthetic Collateralized Debt Obligations (SCDO), Credit Default Swaps (CDS), Credit Default Swap Index (CDSI), Loan only Credit Default Swaps (LCDS), Credit Default Swaps of ABS (ABCDS), Variance Swaps (VS), Constant Proportion Portfolio Insurance (CPPI), Contracts for Difference (CFD)

2. Equity Derivatives: Futures, Asian Options, Barrier Options, Compound Options, Look-back Options, Vanilla Stock Options (put and call options), Vix Options

3. Other Derivatives: Interest Rate Swaps

Both investment strategies assume the application of modern Risk Management practices to mitigate the risk. In general, the risk management is based on the principles of diversification, hedging and risk measurements. The actual risk is measured using the concepts of the Economic Capital and the Credit Modeling:

1.  Cost of Capital is calculated using the Weighted Average Cost of Capital (WACC) model, which includes the following financial variables and ratios: Levered Beta, Debt/Total Capitalization, Tax Rate, Unlevered Beta, Targeted Capital Structure, Risk Free Rate, Market Risk Premium, Spread over Risk Free Rate.

2.  Cost of Equity is calculated using the Capital Asset Pricing Model (CAPM), which includes the following financial variables and ratios: Beta = Firm Specific Risk / Market Risk, Cost of Equity = Risk Free Rate + Beta, Multifactor Models of Asset Returns. In CAPM theory, beta is a measure of risk: a measure of stock price volatility relative to the overall benchmark market index. Beta changes from 0 to 2 (beta=0, risk=0; beta=1, then risk=average market risk (a stock moves up or down in the same proportion as the overall market); beta=2, then risk=well above average market risk).

3.  Monte Carlo computer simulation techniques are used to generate scenarios, and statistical tools to analyze the results.

The four main categories of risks, considered by banks, are in Bernanke (2009):

1. Market Risk;

2. Credit Risk;

3. Operational Risk;

4. Rollover Risk

Other categories of risks may include the transaction risk, foreign exchange risk, reputation risk, emerging markets risk, environmental risk, geopolitical risk, etc. The quantitative techniques such as the Option Pricing, Delta Hedging and Value at Risk are widespread tools for the risk management within financial institutions. Evaluation of the risk of investment projects is based on the concept of calculation of the Net Present Value (NPV). The Autoregressive Conditional Heteroskedasticity model was proposed for statistical modeling of volatility in Engle (2007). However, these techniques could induce similar trading patterns among banks exposed to external influences, and increase systemic risk of multiple failures of financial institutions Barber (2008). Some economists argued that the financial institutions will never have a perfect risk management model in Greenspan (2008, 2007). The reason for this way of thinking is grounded on the perceptions that the risk is in Kay (2009):

1) Subjective expected utility;

2) Uncertainty.

In our opinion, the risk is different from the uncertainty, which can not be controlled or measured, during the risk management at financial institutions in Kay (2009).

In 2006, the Basel Committee for Banking Supervision introduced the Basel II accord, which is a risk-based capital adequacy project with clearly specified minimum regulatory capital requirements for credit risk management by banks in Basel Committee on Banking Supervision (2006).

The financial system is an integral part of market economy. Having discussed the definition of terms and modern practices in finances, let us concentrate on the theoretical definition of business cycle in market economy, and then, think about: How can the business cycle impact the financial system? The time dependence of real Gross Domestic Product (GDP) usually consists of the fluctuations under the long term growth period. The expression $\Delta G(i) = \Delta G(i) - \Delta G(i-1)$ commonly shows the cycle, which is repeating depression and prosperity, is called ''business cycle'' in Taniguchi, Bando, Nakayama (2008). Such business cycle may be characterized into several types according to its period: the 3 – 7 year Kitchin inventory cycle, the 7 –11 year Juglar fixed investment cycle, the 15 – 25 year Kuznets infrastructural investment cycle in Kuznets (1973) and the 45 – 60 Kondratieff long wave cycle in Kondratieff (1935). The origin of any business cycle can be caused by many external and internal sources. Business cycles cause significant variations of economic variables and indices.

Now, let us consider the present state of matters in finances, analyze the reasons of functional failure of financial system and drive conclusions to create a conceptual framework for policymakers and regulators to shape the financial industry in a century of globalization.

The present business cycle is characterized by severe crisis in finances, because of a number of reasons. First of all, it is a systemic crisis, which makes it so different from the previous recessions. The root of crisis is a complete failure of essentially wrong ideas and invalid assumptions on how to govern the global economy and finances in conditions of capitalism, which took over the minds of policymakers. The prevailing opinion was that in Begg (1982) and in Skidelsky (2009):

1) Markets are efficient;

2) Shares are priced correctly;

3) Demand and Supply of goods and services are well balanced by market itself.

These ideas and assumptions provided a foundation for the so-called Monetarism theory, which was developed and supported by monetarists from Chicago school of economics in Friedman (1967), Stigler (1971), Hayek (1980), and had a primary influence on Anglo-American economic and financial policies since 1970s in Skidelsky (2009). The Austrian school of economic thought, founded by Carl Menger, Eugen von Böhm-Bawerk and Ludwig von Mises and further developed by Henry Hazlitt, Friedrich Hayek and Murray Rothbard, who advocated for the strict enforcement of voluntary contractual agreements and inadmissibility of application of coercive forces on transactions between free market agents in Menger (1871), von Böhm-Bawerk (1884–1921), von Mises (1940–1949), Hazlitt (1946), Hayek (1948), Rothbard (1962), became a main source of inspiration for the Monetarism theory advocates. However, the Neoclassical Price Theory and free market libertarianism postulates are very controversial and some economists argued about their validity, for example the idea that "the markets are self correcting" is claimed to be a misperception in Krugman (2009). The free market forces failed in Clarke (2009):

1. To deliver the goods;

2. To offer self-correction;

3. To cope with self-inflicted crisis of confidence.

The prevailing opinion was also strongly criticized in Soros (2008):

1. Paradigm that market tends to an equilibrium is false;
2. Deviations from equilibrium occur in random nature;
3. Assumption that people act rationally is not a case;
4. Calculation of (financial) risk was based on the false paradigm.

The Austrian school of economic thought came up with the Austrian business cycle theory, which views the business / credit cycles as a result of unwise central bank's financial policies execution, which may cause the interest rates to remain very low for long time, resulting in excessive credit creation and expansion, misallocation of capital, law savings accumulation, and finally economic bubble generation in von Mises L (1912), Hayek F A (1931). In the Austrian economists' opinion, the centralized government intervention with the help of central bank in processes of capital allocation in conditions of efficient market results in the business cycles generation. In Danny Quah's words: "Hayek viewed business cycles as having their initiating impulse of central bank credit overexpansion and their propagation mechanism of misallocation of capital across short- and long-term investments" in Quah (2007). The Austrian business cycle theory was strongly criticized by John Maynard Keynes in 1930s, and then it was regarded as incorrect business cycle theory and rejected in Friedman (1969, 1993), Krugman (1998).

Similar theory of business cycles origin, for instance as a result of speculative increases in the value of land, was proposed in George (1879). The introduction of wise taxation policies due to the shift of taxation burden from the capital and labor to the land value was proposed to be considered as an opportunity to avoid the long term recession cycles under free market capitalism in George (1879).

The US Federal Reserve under former Fed's Chairman Alan Greenspan in 1987-2006 set very low interest rates for Federal Reserve funds over extended time period, which, in combination with quite liberal monetary base management policies, made the speculative increases in the value of structured investment vehicles possible, and to some degree, contributed to the severe recession cycle in the USA, and provided the grounds to validate the Hayek's theory assumptions.

In conditions of economic recession cycle, triggered by market imbalances, there is a serious solvency problem among the market agents, because the private sector incomes were not able to continue to service private sector debt in Skene (2009). It was soon realized that this is a most severe financial crisis since World War II in Hutton,

Wolf (2008). The growing inequalities of opportunity and income between the free market agents led to the social unrests in many countries around the World. The present economic problems began to pose a considerable treats to the developed nations in Gamble, Hutton, Quah (2009). The current situation is similar to the time of Great Depression in the USA in Bernanke (2004). Main conclusions on origins of the economic downturn were derived in Soros (2009):

1. Economic cycle has to collapse, because it was built on false premises;

2. The bigger bust, the bigger collapse;

3. During a crash you have a liquidation of mispriced assets.

Most influential economists, financiers and political philosophers were in search for effective means to prevent the business cycles. The common modern view on possible resolution of presently existing problems in finances and economics is that the Adam Smith capitalism model, who believed that the free market's is able to establish a price that provides a fair return on land, labor, and capital as well as to procure the right amount and variety of goods and services by a so-called "invisible hand" in Smith (1776), has to be complemented by the John Maynard Keynes capitalism model in Keynes (1936). The recent action strategies by policymakers around the World are in good agreement with the John Maynard Keynes conjectures about the modern capitalism. The Keynes theory postulates and findings were presented in his well known research paper titled: "The General Theory of Employment, Interest and Money" in Keynes (1936), and are summarized in Skidelsky (2009):

1. We don't know nearly as much about the future as we think;

2. When markets suffer big shocks, they don't sell correct quickly, but start shrinking;

3. Government must inject extra liquidity into economy to compensate for the decline in private spending;

4. Market system is needed central management, if it is to work for everyone's benefit.

Let us consider the operation of global financial system, which is affected by two main problems in Rodrik (2009):

1) Weak financial regulation and supervision;

2) Growth of global economic imbalances.

We begin with the explanation that the central banks in different countries govern the state finances through the creation of financial policies for effective systemic financial regulation. Also, the monetary base is created by central banks in Krugman (2009). The core purposes of central bank is to establish and maintain:

1. Monetary stability, which means stable prices and confidence in the currency;

2. Financial stability, which entails for detecting and reducing threats to the financial system as a whole.

The debate on the encountered problems in financial system and possible remedies to fix the global financial system was initiated at all levels in global society as the financial crisis started to progress in conditions of global recession. The quick action to defuse the financial crisis was needed in Hutton, Wolf (2008). It was agreed that the monetary stability and financial stability have to be better maintained by the central banks. In addition, the critical issues to think about were formulated in Pandit (2009):

1. The need for regulatory structure that will allow markets to clear up efficiently;

2. The need for financial architecture that can truly optimize the global GDP growth;

3. The need for a lot of global coordination.

Let us concentrate on the two main ideas, which were proposed by many researchers and deserve particular attention:

1. **To strengthen the regulation in the finances** and;

2. **To add liquidity in the banking system**.

The **first idea** is focused on the creation of effective systemic regulation for global financial industry. It was agreed in principle that the reputation and functional performance of banks have to be improved significantly. The banks have to become more transparent, demonstrating that their assets are of acceptable quality and offer sufficient margins to compensate for risk. The creation of effective systemic regulation for global financial industry considered the following initiatives in Blankfein (2009):

1) To establish the fair market value for bank's assets;

2) To use an adequate pricing model;

3) To adopt the fair value accounting with responsible systemic risk management;

4) To raise capital requirements for banks with the goal to reduce systemic risk;

5) To improve the underwriting standards;

6) To take to the account all the risks and uncertainties during portfolio risks management;

7) To maintain the culture of transparency and integrity in financial institutions.

In addition, it was argued that some other regulation related measures have to be undertaken to increase efficiency of banking system:

1) The introduction of different sets of regulation for the commercial banking and investment banking may have sense in Peyer (2009).

2) The clear understanding of existing limitations of the Black–Scholes options pricing model, which postulates that the option is implicitly priced if the stock is traded by risk managers in investment banks is essential for the proper Long-Term Capital Management (LTCM) in Black, Scholes (1973) and in Merton (1973). These limitations may include: the tail risk, liquidity risk, volatility risk, gap risk; and have to be considered from position of the Collective Risk Theory in Gerber (1973) and Gerber, Shiu (1994, 2003). Mispricing of value of derivatives resulted in complete bankruptcy of Long-Term Capital Management by investment banks in Gamble, Hutton, Quah (2009). In addition, the understanding of limitations of Modern Portfolio Theory with Efficient Portfolio definition, which is one where no added diversification can lower the portfolio's risk for a given return expectation, will certainly be helpful in Markowitz (1952, 1959).

3) The new regulation for credit-rating agencies, which were not able to perform their main function such as to rate the credit or price the assets of financial institutions correctly, and mistakenly rated many tens of thousands of structured investment vehicles at AAA, has to be developed in Gamble, Hutton, Quah (2009) and in Kay (2009). The very essence of credit rating idea has to be well understood in Becker, Milbourn (2009): "A credit rating is an assessment of the creditworthiness of a corporation or security, based on the issuer's quality of assets, its existing liabilities, its borrowing and repayment history and its overall business performance. Ratings predict the likelihood of default on financial obligations and the expected repayment in the event of default. By making information about these important factors widely available, ratings fulfill a key function of information transmission in financial markets."

4) More regulation of hedge funds has to be introduced in FSA (2009) and in Soros (2009).

5) The reach of clear agreement on banker's compensation packages, bonuses and incentives may certainly be considered as a valuable contribution to the creation of effective systemic regulation for global financial industry in Gamble, Hutton, Quah (2009) and in Roberts (2009).

6) A more in depth understanding of the implications of increased financial interdependence in a globalized world has to be developed and reflected in a new banking regulation in Aznar (2009).

Reviewing all the ideas on the new proposed regulation initiatives, it is necessary to understand that: "What is needed is not more regulation, but better regulation," in Knight (2009).

The **second idea** was based on the common perception that there is a strong necessity to create a new monetary policy, to maintain the monetary and financial stabilities by adding the liquidity to the banking system with the aim to avoid a failure of confidence. The idea was to get the bad debt out of financial system and re-capitalize the banks. This policy of adding liquidity to the financial system was called the Quantitative Easing or Credit Easing in Bernanke (2009), and it was well supported by a number of financiers, who emphasized that the collapse of credit basically means that authorities have in Soros (2009):

1) To re-inflate the system;

2) To arrest the collapse of credit;

3) To re-constitute or re-organize the system.

During the realization of Credit Easing, the Federal Reserve focused on three strategies in Bernanke (2009):

1) To provide short term liquidity to financial institutions;

2) To serve as a source of liquidity for financial institutions in medium term;

3) To purchase the long term securities for Federal Reserve portfolio.

We agree with the statement that the financial markets can be characterized as open, non-linear and complex systems in Beinhocker (2006). **We think that the nonlinear phenomena in the economy have to be taken to the account during the stimulus packages introduction by every central bank.** In our opinion, the fluctuations of magnitudes of economic variables may become large enough to produce substantial nonlinear distortions during the phase transitions caused by:

1) Transition of economy from one business cycle to another business cycle, which may be characterized by critical point, at which the actual transition takes place;

2) Coupling between the business cycles: the Kitchin inventory cycle, Juglar fixed investment cycle, Kuznets infrastructural investment cycle and the Kondratieff long wave cycle.

Let us explain our proposal in details: Every business cycle has a periodic nature, hence it can be represented as a wave with certain frequency of oscillation, time period and amplitude of oscillation in accordance with the Theory of Electromagnetic Field in Maxwell (1861, 1865 and 1873). **We think that the mixing of business cycles during their interaction may lead to an appearance of the frequency, phase and amplitude modulation phenomenon, which, under certain conditions, may result in origination of strong nonlinear dynamics in financial and economic systems accompanied by the intermodulation distortions, phase noise and chaos.** For example, the mixing of the Kitchin inventory cycle short wave, Juglar fixed investment cycle short wave, Kuznets infrastructural investment cycle medium wave with Kondratieff long wave may produce, under certain conditions, the intermodulation distortions resulting in nonlinear dynamics in global financial system. For instance, the random shift of phase of modulated wave during the mixing of Kuznets infrastructural investment cycle medium wave with Kondratieff long wave may generate the phase noise. In addition, the changing degree of entropy of financial/economic system, which is defined by financial/economic variables and ratios described early, may generate the different kinds of nonlinearities and chaos in finances in Mosekilde (1996). Therefore, the nonlinear dynamics has to be taken to account during both the probabilities computing in risk management methodologies in finances and the introduction of stimulus packages by central banks into financial system, which, in some cases, may even be affected by the so called Butterfly effect: sensitive dependence of financial system on initial conditions in Lorenz (1963, 1964) and Gleick (1987).

**We propose the idea that, during the implementation of Quantitative Easing policy by the Federal Reserve in the US or by Bank of England in the UK, the liquidity has to be added to the financial system by small quantas (in small quantities, parts or steps) in series over time period**, in analogy with quantum physics in Feynman (1962, 1965). Every time, the decision about the adding of extra quanta of liquidity to the financial system by central bank must be based on the analysis of obtained product of reaction of negative/positive feedback loop, existing in financial system, during the extended periods of market disequilibrium. The expectation to solve the crisis in finances with the addition of big extra liquidity packages to the financial system promptly, as it was done by Federal Reserve and approved by Congress in the USA in 2009, may lead to the inefficient allocation of capital among the banks, which may pursue the short term investment strategies to invest all the extra liquidity resources in the Structured Investment Vehicles or to speculate on the foreign exchange money markets with the aim to get extra profit from speculation. These big profits, made by some of Wall Street's leading banks, are

labeled as the "hidden gifts" from the state in Soros (2009) and in Freeland (2009). Therefore, we suggest encouraging banks to pursue the long term investment strategies in productive capacities. It can be done with the introduction of a new tax. **We propose to introduce the Random Tax, which must be imposed on the profits obtained by market agents as a result of their random decisions to pursue and complete the high-risk high-profit transactions, which, as a side effect, tend to increase the volatility and destabilize the free market.** The basic idea is that the Tax authorities will encourage and/or coerce all the market agents (banks, investment firms and hedge funds) to act responsibly and creatively rather than irresponsibly and destructively, when conducting the transactions in the free market, by selecting the market agents, which engaged in high profit speculative transactions (these events are random) and by imposing the Random Tax on the profits obtained during these transactions. The Random Tax will certainly complement the present taxation regulation, stabilize the free market, improve the generally accepted free market rules and turn the free market to a better place to make business.

Similar idea was to introduce a global tax on financial transactions: "Tobin tax" in Tobin (1971, 2001). Tobin tax involves applying a small charge (0.1%) on foreign currency transactions to protect countries from exchange-rate volatility caused by short-term currency speculation in Pimlott (2009) and in Tobin (2001). The Tobin tax was supported by many leaders in Brown (2009) and in Giles (2009). Other proposed options to consider with the aim of encouraging of responsible banking may include in Pimlott (2009):

1) Tobin tax a global financial transactions levy;

2) Insurance fee to reflect systemic risk;

3) Debt that swaps into equity when capital levels fall too low;

4) Resolution fund, similar to a pre-funded bank bail-out fund, that lenders would pay into and could be used should an institution collapse.

In our opinion, the Random Tax is better than the Tobin Tax, because it is a tax, which applies to the high-risk high-profit speculative transactions between the market agents only. The Random Tax may be imposed selectively, based on the generally accepted criteria and definitions of this type of transactions.

The next macroeconomic problem is that the excessive liquidity in financial system may help to fight with the deflation in financial system in short term, but result in an increase of inflation expectations in long term. In the USA, the money became a commodity rather than a medium of exchange in time, when the US dollar enjoys a privilege status of global reserve currency. Foreign investors prefer to invest the earned US dollars in the US government bonds with 2%-3% interest rate guaranteed by the state. In near term perspective, the existing interest rates on the US government bonds may become smaller than the inflation expectations, caused by the US Quantitative Easing policies, soaring US trade deficit, collapsed US real estate market and US economic downturn. Main creditors of the USA: P.R. China, Japan and gulf countries, may prefer to diversify their monetary reserves with the aim to hedge against the weakening US dollar and systemic risks associated with the US financial system. Therefore, the wise exit strategy from the Quantitative Easing and control of inflation is a task number one for the US Federal Reserve and US Government.

The growing trade deficit in the USA; big trade surplus in P.R. China; devaluation and fixation of Chinese currency at undervalued level lead to asymmetric capital flows between the financial markets. These global imbalances can be partly controlled with the help of multilateral currencies exchange rates agreement in Flassbeck (2009).

The financial system has to be efficient. It means that the size of banks and structure of the national financial systems have to be changed. The small and medium size local banks are more efficient in serving the needs of small and medium size businesses, because their managerial systems as well as capital supply and distribution chains are more optimized for this purpose in Justin Yifu Lin (2009). The small banks can react promptly on the constantly changing local market conditions. Therefore, the capital intensive large inefficient banks and investment firms with high degree of bureaucracy have to be re-shaped in accordance with the new financial policies framework and specific needs of every country. However, the idea to break up the banks and investment firms with big capitalization has to be approached carefully, because the role of large systemic banks and investment firms in supply of capital for the transnational corporations or governmental housing/education loans national programs shouldn't be underestimated.

In addition, the new capital requirements and acceptable risk management practices have to be introduced and maintained by every bank. All the banks will need in Sachs (2009):

1) To have the depositary reserve;

2) To maintain the capital adequacy;

3) To have the depositary insurance.

The Basel Committee for Banking Supervision has to develop the new international banking standards with revised:

1) Bank capital rules;

2) Leverage ratios rules;

3) Liquidity rules.

Five wise advices are given to financial authorities in Soros (2009):

1) To accept responsibility for preventing bubbles from growing too big;

2) To control the money supply; to control the availability of credit by setting the margin requirements and minimum capital requirements;

3) To monitor the systemic risks;

4) To recognize that financial markets evolve in a one-directional, non-reversible manner, and extend an implicit guarantee to all financial institutions that are too big to fail;

5) To raise the risk ratings of securities held by banks similar to the risk ratings of regular loans in the next Basel Accords.

We are confident that the financial innovation will continue to generate the wealth in long term perspective. The creation of new financial clusters will enable this innovation, because the clusters will encourage competition among the banks in Porter (2008).

We think that the global financial meltdown had a negative impact on the reputation of the international banking system. This paper has aimed to propose the conceptual intellectual framework on new banking regulation policies, which may be used by G20 policymakers to regain confidence in financial system through the creation of new financial architecture in a century of globalization. The G20 policymakers and regulators may consider introducing the proposed changes into the global financial system, aligning it with principles of transparency and integrity with the aim to establish a truly competitive environment for sustainable accountable prosperous development of international banking system. This open environment will certainly facilitate the realization of competitive strategies toward the effective, profitable, responsible and sustainable bank operation with the focus on in Chaney, Goodhart, Webb (2009):

1. Governance of banks;

2. Fair valuation of assets;

4. Optimization of risk management;

5. Creation of effective compensation system.

We believe that the US Federal Reserve, European Central Bank, National Central Banks, Basel Committee for Banking Supervision, International Monetary Fund, World Bank, Bank for International Settlements and Financial Services Authority will effectively collaborate together to establish the reputation of international financial system with introduction and adaptation of better financial regulation. We are confident that the globalization would work as in Wolf (2009), if the improved global governance policies, created due to our advanced mental understanding of universal balancing economic laws in Brochmann (1929, 1930, 1931), could better balance the global capital, commodity, equity and labor markets on the way to the financial stability and economic prosperity.